# Albert Einstein's close friends and colleagues from the Patent Office

Galina Weinstein

In the Patent Office Einstein hatched his most beautiful ideas, and there he spent his "Happy Bern Years". These wonderful ideas led to his miraculous year works of 1905. Einstein was not an expert in academic matters, and he was out of academic world. Neither did he meet influential professors, or attend academic meetings. He discussed his ideas with his close friends and colleagues from the Patent Office. In 1907 he finally got his foot into the academic doorway; Einstein became a privatdozent and gave lectures at the University of Bern. However, his first students consisted again of his two close friends and another colleague from the Patent Office.

## 1 Physics Group

### 1.1 The Patent Office

On June 23$^{rd}$, 1902 Einstein started his new job as a technical expert (provisional) third-class in the Swiss Patent Office, with a salary of 3500 francs per year. The Patent Office was on the upper third-floor of the new building of the Postal and Telegraph Administration, near the railroad station and the medieval Clock Tower in Bern.[1]

Einstein's first home in Bern was a small room in Gerechtigkeitsgasse 32 (from Feb 11, 1902 until May 31, 1902).[2] From this one-bedroom flat he walked every morning to the building of the Postal and Telegraph building. One day Max Talmud (Talmey) visited Einstein in Bern and saw his flat: "In April 1902, […] I found my friend there and spent a day with him. His environment betrayed a good deal of poverty. He lived in a small, poorly furnished room. I learned that he had a hard life struggle with the scant salary of an official at the Patent Office. His hardships were aggravated through obstacles laid in his way by people who were jealous of him. As a token my friend gave me a reprint of his first scientific publication. It had appeared shortly before in

---

[1] Frank, Philip, *Einstein: His Life and Times*, 1947, New York: Knopf, 2002, London: Jonathan, Cape, p. 23; Frank, Philip, *Albert Einstein sein Leben und seine Zeit*, 1949/1979, Braunschweig: F. Vieweg, p. 45.
[2] Flückiger, Max (1960/1974), *Albert Einstein in Bern*, 1974, (Switzerland: Verlag Paul Haupt Bern), p. 134.
Einstein's second home was at Thunstraße 43a from June 1 1902 until August 14 1902, and his third place was on Archivstraße 8 from August 15, 1902 until October 15, 1903, and then before he got married he moved again to Tillierstraße 18 on October 16 and left on October 28, 1903.

the *Annalen der Physic* (1901, 4 F) under the title […] (Influence of the Capillarity Phenomena)."[3]

Einstein's work at the patent office involved examining the submitted patent applications. The Patent Office received the application and had to legally protect the inventor and the invention against infringement. Therefore a patent examiner was required to have knowledge of patent law. In addition he had to reformulate the patent application in an accepted statement and clear and logical language that defined and explained the invention. Thus he had to be able to read technical specifications.[4]

Einstein's sister, Maja, explained it further in her biographical sketch of Einstein. She said that Einstein's job at the Patent office was on the whole interesting, because it included insight into inventions and registering patents. Maja described the job routine: the activity was to follow the patent law, to see that nothing in the patent application was wrong. The assignments were limited with the intention of formulating in a technical, legal, logical and linguistic language the patent application. One could never know what could be invented in those days, impossible things or else possible working things and possible special arrangements, but sometimes described by the inventors in an awkward manner or even in funny terms. Einstein was supposed to formulate the application in a clear language.[5]

Since under the rules, all applications for patents were destroyed after eighteen years of patent protection we cannot know exact details about the specific patents that Einstein examined. Even in the 1920s, when it was realized that no other employee of the Bern Patent Office or any patent office anywhere, would ever rise to Einstein's heights, and Einstein was world-famous, neither Friedrich Haller nor his successor wished to make an exception from that rule for the benefit of future biographers, says Einstein's biographer, Albrecht Fölsing. Therefore Einstein's comments on inventions were disposed of until 1927. [6]

Only one of Einstein's expert *comments* survived, because it found its way into court records and survived there. It was compiled on December 11, 1907, and rejected a patent claim by the AEG Company of Berlin for an alternating-current machine as "incorrectly, imprecisely, and not clearly prepared".[7]

---

[3] Talmey, Max, *The Relativity Theory Simplified: And the Formative Period of Its Inventor*, 1932, New York: Falcon Press, Darwin Press, p. 167.
[4] Frank, 1947/2002, p. 23; Frank, 1949/1979, pp. 45-46.
[5] Winteler-Einstein Maja, *Albert Einstein –Beitrag für sein Lebensbild*, Einstein Archives, Jerusalem: the full biography printed in a typing machine with a few missing pages and double pages, 1924, p. 21.
[6] Fölsing, Albrecht, *Eine Biographie Albert Einstein*, 1993, Frankfurt am Main: Suhrkamp. English abridged translation by Ewald Osers: *Albert Einstein, A Biography*, 1997, New York: Penguin books, p. 104.
[7] Swiss Patent Office Letter on the AEG Alternating Current Machine, *The Collected Papers of Albert Einstein. Vol. 5: The Swiss Years:Correspondence, 1902–1914* (*CPAE*, Vol. 5), Klein, Martin J., Kox, A.J., and Schulmann, Robert (eds.), Princeton: Princeton University Press, 1993, Doc. 67.

When Einstein was asked how things functioned in the Patent Office, he replied that above all one must be able to express clearly and correctly the wording of the original patent from the description of the discovery and the patentee's claims. The work was not particularly exciting and apart from one or two exceptions it was rather soul-destroying. In any case one had to sit every day for eight hours on a stool and in return for that one was given a decent wage.[8]

By 1905 Einstein may not have been an expert in academic matters, but it appears that he was an expert in the work of the Patent office. Einstein had little knowledge of the latest scientific publications, because he was out of academic world and worked eight hours a day in the office. Einstein probably read papers in the *Annalen der Physik*, *Physikalische Zeitschrift*, and other German physics journals. Einstein's knowledge was probably not updated as it would have been had he stayed in some university, or met physicists in more organized meetings.

Such meetings were the best opportunity for meeting influential professors and even finding position in the academic job market. Physicists then would meet in specialized groups within the framework of the annual general meeting of the Deutsche Gesellschaft der Naturforscher und Ärrzte, the German Society of Scientists and Physicians. In 1906 the society met in Stuttgart and Einstein could have become acquainted with the leaders of his field.

*Physikalische Zeitschrift* had put beforehand a notice.[9] However, Einstein was absent, probably because of his duties to the Patent Office. The following year, when the meeting was held in Dresden, Einstein was again absent.

Einstein got some of his best ideas of Patent office, as he later told his best friend Besso on December 12, 1919, "I was very interested that you want to go again to the Patent Office, in this weltliche Kloster [worldly cloister] I hatched my most beautiful thoughts, and there we spent such happy days together. Since then, our children have grown up and we have grown old boy!"[10] Einstein liked describing the Patent Office to his friends as a worldly cloister.[11] And he used to call his life in Bern: "Happy Bern Years".[12] Einstein was, in any case, not a professional type at all, and he valued his independence more than any formal position".[13] This impression of Dr. János Plesch from around 1919 perfectly represented Einstein's patent years in Bern.

---

[8] Seelig Carl, *Albert Einstein: A documentary biography*, Translated to English by Mervyn Savill 1956, London: Staples Press, p. 55; Seelig Carl, *Albert Einstein; eine dokumentarische Biographie*, 1954, Zürich: Europa Verlag, p. 67.
[9] "Tagesereignisse. An läßlich der im September dieses Jahres im Stuttgart Stattfinden 78. Versammlung Deutscher und Ärzte wird, wie in den früheren Jahren, eine Ausstellung naturwissenschaftlicher und medizinscher Gegenstände stattfinden, welche auf Neuheiten beschränkt sein soll". *Physikalische Zeitschrift* 7, 1906, p. 432.
[10] Einstein to Besso, December 12, 1919, Einstein, Albert and Besso, Michele, *Correspondence 1903-1955* translated by Pierre Speziali, 1971, Paris: Hermann, Letter 51.
[11] Seelig, 1956, p. 56; Seelig, 1954, p. 68.
[12] Die "glücklichen Berner Jahre"; Herneck, Friedrich, *Albert Einstein: ein Leben für Wahrheit, Menschlichkeit und Frieden*, 1963, Berlin: BuchVerlag der Morgen, p. 77.

Einstein felt free; free of academic authority and rules, and thus the Patent Office was a comfortable place for Einstein the free-thinker to hatch his most beautiful theories. Rudolf Kayser, Einstein's son-in-law writes in his biography on Einstein, "He soon discovered that he could find time to devote to his own scientific studies if he did his work in less time. But discretion was necessary, for though authorities may find slow work satisfactory, the saving of time for personal pursuits is officially forbidden. Worried, Einstein saw to it that the small sheets of paper on which he wrote and figured vanished into his desk-drawer as soon as he heard footsteps approaching behind his door. If he had been discovered, he would have been ridiculed as well as harmed. The Director [Friedrich Haller] would have laughed at him in addition to being angry; he was too great a positivist to think much of speculative science".[14]

It appears that Einstein used to write his notes on "small sheets of papers", because later one of the students at the University of Zurich described him entering class with notes "the size of a visiting card on which he had scribbled what he wanted to tell us".[15] On these papers Einstein very likely wrote his path breaking 1905 papers; and these small sheets of papers the size of visiting cards could perfectly enter his desk-drawer without being discovered by Haller.

Fölsing writes that, the German Physicist Rudolf Ladenburg told his physics student of his visit to Bern. He told him that he there saw Einstein pulling out a drawer in his desk and announcing that this was his department of "Theoretische Physik", theoretical physics.[16] Whether Ladenburg really visited Einstein in the Patent office before 1905 it is doubtful, he probably did not visit him in Bern at that time.[17] For Biographers, Besso (with Joseph Sauter) and Einstein's table and drawer, and especially the thoughts that Einstein had hatched in the Patent Office, eventually turned to be the best department of Theoretische Physik in the world.

## 1.2 Michele Besso, Joseph Sauter, and Lucian Chavan

Although Einstein did not meet influential professors, he did discuss his ideas with his friends in the Patent Office. Toward the end of 1896 or the beginning of 1897, during Einstein's first semester in the Polytechnic, he had met Michele Besso at the Zürich home of a woman named Selina Caprotti, where people would meet to make music on Saturday afternoons.[18] Einstein and Besso were best friends. Toward the end of 1903 a vacancy for a "technical expert II class" examiner in the Patent Office was advertised to which Einstein immediately drew Besso's attention, the latter joined Einstein on March 4 as a member of the Patent Office. The two, Einstein and Besso, therefore could go from home to the Patent Office and back together.[19]

---

[13] Plesch, János, *János. Ein Arzt erzählt sein Leben*, 1949, Paul List Verlag, München / Leipzig (first published 1947); Plesch, John, *János, The Story of a Doctor*, translated by Edward Fitzgerald, 1949, New York, A.A. WYN, INC, p. 202.
[14] Reiser, Anton, *Albert Einstein: A Biographical Portrait*, 1930, New York: Dover, p. 66.
[15] Seelig, 1956, p. 100; Seelig, 1954, pp. 119-120.
[16] Fölsing, 1993/1997, p. 222; 1993, p. 254.
[17] Seelig, 1956, p. 85, Seelig, 1954, p. 100; Ladenburg was among the early physicists to recognize Einstein's theory of relativity, and to respond to it. He was at the time a German research center at Breslau, and later moved to the Kaiser Wilhelm Institute in Dahlem (where Einstein later headed the physics institute), and then moved to Princeton.

While in the Patent Office, one colleague whom Einstein was not particularly intimate with was Dr. Joseph Sauter, eight years older than Einstein. Einstein respected him very much but they were not close friends. Sauter was a French Swiss, who had studied at the Polytechnic, had been active as chief assistant to Professor Heinrich Friedrich Weber, and served from 1898 to 1936 as technical expert in the Patent Office.[20]

Einstein's third friend in Bern was Lucian Chavan. Chavan came from Nyon in Lake Geneva, and was eleven years older than Einstein. Since 1900 he had been appointed technical secretary in the Federal Postal and Telegraph Administration in Bern.[21]

In August 16, 1952 Carl Seelig wrote to Einstein that by a happy coincidence the librarian of the Federal Postal and Telegraph Administration in Bern in western Switzerland, kindly let him look at the manuscript collection of notebooks dispensed by Einstein's friend from the Patent Office, Lucian Chavan. Chavan died at the age of 74. The notebooks deposited in the library after Chavan's death in August 1942 are the intellectual fruits of his private studies and scientific lectures with Einstein. In these neatly written Notebooks Seelig found photographs and newspaper cuttings of Einstein, all glued (very likely also in a magnificent order).[22]

Under one of these photos (a profile of Einstein from the patent office), Chavan had written: "Einstein is 1,76 meter high, broad shoulders and slight stoop; unusually broad short skull. Complexion a matt light brown. A garish black moustache sprouts above his large, sensual mouth. Nose rather aquiline, and soft deep dark brown eyes.

LUCIEN CHAVAN

Der nach seiner Demission in Bern (1921) nach Genf zog, ausgeliehen. Sie spulen in diesen säuberlich geführten, rührend bildungshungrigen heften eine beträchtliche Rolle. Es sind auch Fotos und Zeitungsausschnitte von Ihnen eingeklebt. Unter eine hat Chavan eine Art "Steckbrief" geschrieben ( 1,76 Mt. gross, sinnlicher Mund, sehr braune Augen... Einstein parle fort correctement le français, avec un léger accent étranger"… Item 39 029, Einstein Archives.

The voice is compelling, vibrant like the tone of a cello. Einstein speaks correct French with a slightly foreign accent".[23]

Seelig perhaps heard that Einstein was late to talk. In 1930 Reiser wrote "Slowly, and only after much difficulty, he learned to talk. His parents thought he was abnormal".[24] Seelig then asked Einstein, "and the legend was spread that you don't have a talent for languages. I admire you even more now, because I have to hassle with even the Germans! Where have you just learnt French so well?"[25]

On 25.8.52, Einstein replied to Seelig: "My French has never been very good".[26] And in another letter from summer 1952, he explained to him what he meant by "French has never been very good",[27]

In the summer of 1909 the University of Geneva bestowed over a hundred honorary degrees in celebration of the 350[th] anniversary of its founding by Calvin. Einstein wrote Seelig the following: "One day I received in the Patent Office in Bern a large envelope out of which there came a sheet of distinguished paper. On it, in picturesque type (I even believe it was Latin) was printed something that seemed to me impersonal and of little interest."

Einstein's secretary Helen Dukas said that it was actually French, printed in script letters.[28] Einstein though believed it was Latin, "So right away it was flung into the wastepaper basket". The invitation was indeed picturesque and written *in French*; it had the date on it: 2 Julillet 1909. [29]

Later, Einstein told Seelig that he learnt it was an invitation to the Calvin festivities and an announcement that he was to receive an honorary doctorate from the Geneva University. Einstein's secretary explained why Einstein had flung this sheet of paper to the waste basket: "There was a remarkable misprint in the impressive document, and this may have registered in Einstein's subconscious and influenced his action: The recipient of the degree was given not as 'Monsieur Einstein' but as 'Monsieur Tinstein'". M. Tinstein did not reply to the invitation.[30]

---

[23] Seelig, 1956, p. 58; Seelig, 1954, pp. 70-71.
[24] Reiser, 1930, p. 27.
[25]"[...] und dabei wird die Legende verbreitet, dass Sie keine Sprachbegabung haben. Jch bewundre Sie nun noch mehr, denn ich muss mich sogar mit dem Deutschen herumquälen! Wo haben Sie nur so gut französisch gelernt?". Item 39 029, Einstein Archives.
[26] "Das Französisch war nie sehr gut". ETH Archives, Zurich:
http://www.library.ethz.ch/en/Resources/Digital-collections/Einstein-Online/Princeton-1933-1955.
[27] "Eines Tages erhielt ich im Berner Patentamt ein grosses Couvert, aus dem ein nobles Papier herauskam, auf dem in pikaresker Druck (ich glaube sogar auf Lateinisch) etwas stand das mich unpersönlich und wenig interessant anmute und sofort in den amtlichen Papierkorb flog. Später erfuhr ich, dass dies eine Einladung zur Calvinfeier war nebst Ankündigung, dass ich an der Genfer Universität den Ehrendoktor bekommen sollte. Die Leute dort interpretierten offenbar mein Schweigen richtig und wandten sich an meinen Freund und Schüler, den Genfer Chavan, der in Bern lebte. Diesser überredete mich, nach Genf zu kommen weil dies praktisch unvermeidlich wäre, ohne weitere Erklärung". Item 39 033, Einstein Archives.
[28] Dukas Helen, and Hoffmann, Banesh, Albert Einstein, *The Human Side, New Glimpses from his Archives*, Princeton: Princeton university Press,1979, p. 5.
[29] Item 39 034, Einstein Archives.

After the first part of the story Einstein told Seelig the second part,[31]

The people at the university interpreted his silence correctly and turned to his friend who came from Geneva Lucian Chavan. The later convinced Einstein to come to Geneva, and they both went there without Einstein knowing the purpose of the journey.[32] Finally Einstein found out what was going on, and tried to avoid the festivities because he had with him only a straw hat and his everyday suit. His proposal to stay away was rejected.

And Einstein ended the story with a typical Einstein joke,[33]

"The celebration ended with the most opulent banquet that I have ever attended in all my life"; and one can imagine Einstein sitting with a straw hat and telling his Genevan neighbor the well-known joke reported by Seelig in his biography: "So I said to a Genevan patrician who sat next to me, 'Do you know what Calvin would have done if he were still here?' When he said no and asked what I thought, I said: 'He would have erected a large pyre and had us all burned because of sinful gluttony'. The man uttered not another word, and with this ends my recollection of that memorable celebration".[34]

Seelig included this story in his 1954 biography of Einstein.[35] This anecdote of Einstein with the straw hat was reproduced in many non-documentary biographies of Einstein. While telling this story biographers presented Einstein as a shabby dressing

---

[30] Dukas and Hoffmann, 1979, p. 5.

[31] "So fuhr ich am angesagten Tag ab und traf Abends einige Züricher Professoren im Restaurant des Gasthofes wo wir wohnten. So hockten wir am Abend beisammen; Professor Taschirch, der Pharmakologe, ein Mann mit Humor, war unter ihnen. Jeder von ihnen erzählte nun, in welcher Eigenschaft er da war. Als ich schwieg, erging die Frage an mich und ich musste gestehen, dass ich keine blasse Ahnung habe. Die andern wussten aber Bescheid und weihten mich ein. Am nächsten Tag sollte ich im Festzug marschieren und hatte nur Strohhut und Strassenanzug bei mir. Mein Vorschlag, mich davon zu drücken, wurde mit Entschiedenheit abgelehnt und diese Feier verlief entsprechend drollig was meine Mitwirkung an langste". Item 39 033, Einstein Archives.

[32] Chavan only persuaded a reluctant Einstein to go to the festivities in Geneva, but did not actually go with him, or else escorted him to Geneva and immediately left back to Bern. Einstein wrote Chavan from Geneva on July 9, 1909, "Leiber Herr Chavan & geehrte Frau Chavan! Ich sende Ihnen einen herzlichen Gruss aus dem gastfreundlichen Genf. Ich bin entzückt über die Freundlichkeit & Liebenswürdigkeit, mit der mir die Leute entgegenkommen". Einstein to Chavan, July 9, 1909, *CPAE*, Vol. 5, Doc. 170, note 2, p. 202.

[33] "Das Fest endete mit dem opulenz den Festessen, dem ich in meinem ganzen Leben beigewohnt habe. Da sagte ich zu einem Genfer Patrizier, der neben mir sass: "Wissen Sie Calvin gemacht hätte, wenn er noch da wäre?" Als er verneinte und mich um die Meinung fragte, sagte ich: "Er würde einen grossen Scheiterhaufen errichtet und uns alle wegen sündhafter Schlemmerei verbrannt haben". Das Mann sprach kein Wort mehr und damit endet meine Erinnerung an die denkwürdige Feier". Item 39 033, Einstein Archives.

[34] Dukas and Hoffmann, 1979, pp. 5-6; pp. 119-120; Seelig, 1956, p. 92; Seelig, 1954, pp. 108-109.

[35] Seelig, 1956, p. 92; Seelig, 1954, pp. 108-109.

weird genius, and they thus told a story of some Monsieur Tinstein and not of Albert Einstein.

Einstein was probably playing tricks a little with Seelig in this story because Seelig annoyed him with long personal questions. In fact, this was not the first time that Einstein was fooling people who reported about his personal life.[36]

Einstein did not quite like personal biographical sketches of his life, and this was the best way to deal with personal questions: *his Swabian sense of humor*. This sense of humor of Einstein also led to a most pompous philosophical discussion group.

## 2 Philosophy Group

### 2.1 Maurice Solovine and Conrad Habicht

Immediately on arriving in Bern, in order to earn some money, Einstein had advertised his services in a local paper, the *Berner Stadtanzeiger*, as a private tutor in mathematics and physics, with the added bonus of "free trial lessons". Solovine told the story many years later: [37]

"Walking one day in the streets of Bern during the Easter vacation of 1902, I bought a newspaper and happened to see an announcement saying that Albert Einstein, former student at the Zürich Polytechnic, offered physics lessons for three francs an hour. I said to myself: 'Perhaps this man can introduce me to the mysteries of physics'. I found my way to the address given, went up the stairs and rang the bell. I heard a loud *Herein!*, and then Einstein appeared. As the door of his apartment opened into a dark hallway, I was struck by the extraordinary brilliance of his large eyes.

After entering his room and sitting down, I explained to him that I was studying philosophy, but also wished to deepen my knowledge of physics so as to obtain a firm knowledge of Nature. He told me that he himself, when younger, had had a keen interest in philosophy, […] and he now confirmed himself to physics".

Solovine recalls that their first conversation lasted almost two hours, "about all sorts of questions, and it seemed that we had a community of ideas and a personal affinity", after which Einstein escorted Solovine to the street. "When I prepared to leave, he came with me, and we conversed in the street for about half an hour before agreeing to meet again the next day…".[38] In the third meeting Einstein declared that meeting for free discussions of philosophy would be much more fun than paid lessons. They decided to read together the works of the greatest philosophers and afterwards discuss their ideas. Einstein seemed to have liked Solovine for he later wrote Seelig on 5 May 1952, "der überaus gütige Solovine" (the very kind Solovine).[39] They were soon joined by another friend.

---

[36] In 1919 Einstein a *New York Times* correspondent came to his Berlin home to interview Einstein about the thought experiment of the man falling from the roof he had invented while in the patent office. The reporter reported, "It was from his lofty library, in which this conversation took place [Berlin] that, he observed years ago a man dropping from a neighboring roof – luckily on a pile of soft rubbish – and escaping almost without injury. This man told Dr. Einstein that in falling he experienced no sensation commonly considered as the effect of gravity […]". Presumably Einstein told the correspondent the story in this way, and he did not notice that Einstein was fooling him. "Einstein Expounds his New Theory", *New York Times* 1919, December 3.

[37] Einstein, Albert, *Lettres à Maurice Solovine*, 1956, Paris: Gauthier Villars; *Letters to Solovine, with an Introduction by Maurice Solovine* (With an Introduction by Maurice Solovine, 1987), 1993, New York: Carol Publishing Group, p. VI; Solovine, 1979, p. 9.

"Several weeks later", Solovine recalls, Conrad Habicht also joined the meetings and the three formed a group, a discussion club. Solovine recalls that Einstein met Habicht during his stay in Schaffhausen (Habicht's hometown), the latter came to Bern to complete his (PhD) studies, as he wanted to teach mathematics in secondary school. Habicht was working on a doctoral thesis in mathematics.[40]

Habicht was elected in 1904 to the post of mathematics and physics teacher in the Protestant Educational Institute in Schies (Graubünden) where he stayed until his return to Schaffhausen in 1914.

Einstein's relations with Habicht and Solovine were very close. On January 3, 1903, Einstein got married with Mileva Marić. Mileva was 27 and Albert, 24. The wedding took place at the registry office in the old city; no wedding guests had arrived, either from Einstein's family, nor from Marić's family. The witnesses were Conrad Habicht and Maurice Solovine. In January they moved to Albert's small apartment in the area Kirchenfeld near the Aare River, on Tillierstrasse 18. In November 1903 the Einsteins moved into the city, renting a third-floor apartment at Kramgasse 49. This apartment was very modest, reached by a steep narrow staircase, and consisted of two rooms only, one with large windows looking out the fine street. This was where Einstein's son Hans Albert was born, on May 14, 1904. The Einsteins gave up this apartment on May 16, 1905 and moved to Besenscheuerweg 28 in the Mattenhof district on the outskirts. They moved to be closer to Michele Besso and his wife.[41]

Much later, in May 1905 Einstein wrote Habicht the famous letter announcing his four miraculous year papers: "But why have you not yet sent me your [doctoral] thesis [on mathematics]? Don't you know, you wretch, that I should be one of the few fellows who would read it with interest and pleasure? I can promise you in return four works, the first of which I shall soon be able to send you as I am getting some free copies. […]".[42] Einstein presumably wanted to explain to his friend Habicht his latest path breaking papers, and discuss with him the draft of his paper on relativistic kinematics.

In September 1905 Einstein tried to persuade Habicht to come and work with him in the Patent Office: "if an opportunity arises I shall give you a boost with Haller, perhaps we may manage to smuggle you in among the patent boys […] Would you, in fact, be prepared to come? Think that each day there are eight working hours, leaving eight hours for leisure, and then there is Sunday. I should be very pleased if you were here". Einstein told his friend that he did not have to bother about his valuable time in the patent office, because "there is not always a subtle theme to meditate upon. At least, not an exciting one".[43]

A year and a half later, either on April 27 or on May 3, 1906, Einstein was trying to persuade his other friend, Solovine, to come and work with him in the Patent Office: "On top of my curiosity, a small practical matter has now come up, and that's why I'm writing to you. A few days ago a local patent attorney, to whom I mentioned you in the past, presented me with a document that had to be translated into impeccable French. Of course, I did not take on the task, since the matter was urgent. But I wish to ask you all the same whether you have found some more or less satisfactory means of livelihood. If not, then there still is some possibility that you might find work, and in due course even a permanent appointment, here in a Patent Office. Do write me soon what you think of that".[44]

## 2.2 The Olympian Academy

The three, Solovine, Habicht and Einstein formed a discussion club and decided to poke a little fun at pompous scholarly societies. Maurice Solovine, the philosopher among the participants in the group, thus chose the grandiose name "Akademie Olympia" (Olympia Academy)[45] to their group. Even though Einstein was the youngest of the three, he was designated the president, "Akademiepräsident", and ironically he was called, "Albert Ritter von Steißbein" ("Albert the knight from Backbone").[46] Solovine prepared a certificate with a drawing of Einstein's head in profile,[47]

"The Man of Hechingen,

Expert in the noble arts, versed in all literary forms – leading the age toward learning, a man perfectly and clearly erudite, imbued with exquisite, subtle and elegant knowledge, steeped in the revolutionary science of the cosmos, bursting with knowledge of natural things, a man of the greatest peace of mind and marvelous family virtue, never shrinking from civic duties, the most powerful guide to those fabulous, receptive molecules, infallible high priest of the church of the poor in spirit".

[43] Einstein to Habicht, 30 June-22 September 1905, *CPAE*, Vol. 5, Doc. 28.
[44] Einstein to Solovine, April 27, 1906, *CPAE*, Vol. 5, Doc. 36; Einstein, 1993, Einstein to Solovine, May 3, 1906, pp. 18-19 (The original German letter in Einstein's handwriting was undated, and therefore the date was inferred according to the events in the letter).
[45] "die unsterbliche Akademie Olympia". Herneck, 1963, p. 72.
[46] Herneck, 1963, p. 70.
[47] Maurice Solovine, "Dedication, Einstein as Member of the Olympia Academy", A.D. 1903, *CPAE*, Vol 5, Doc. 3.

Hoffmann and Dukas say that "Einstein was the leading spirit" in the group.[48] The friends read and discussed together major works of philosophy and science that powerfully influenced the development of Einstein's ideas. "The Olympian Academy was in earnest, and above all it was fun".[49]

Solovine recalls that, "Einstein also had us dine together. These dinners were of an exemplary frugality. The menu was usually made up of sausage, a piece of Gruyère cheese, fruit, a small jar of honey and one or two cups of tea. But we were overflowing with good spirits…"[50] They used to pursue their discussions in the Café Bollwerk, a few steps from the Patent Office, and also in their rooms.

One day Solovine said to Habicht, "Let's give Einstein a special surprise, and serve him caviar for his birthday on March 14[th]." Solovine then recalls: "On 14 March we went to dinner at his apartment and, in just the same way as I would have served the sausage, etc, I put the caviar on three plates on the table before joining Einstein. As chance would have it, he was talking that evening about Galileo's principle of inertia, and in so doing lost all consciousness of worldly joys and tribulations. When we sat down at the table, Einstein took mouthful after mouthful of caviar, while continuing to talk about the principle of inertia. Habicht and I furtively exchanged astonished glances, and when Einstein had finished all the caviar I said: 'Do you realize what you have been eating?' Looking at me with his big eyes he said: 'What was it, then?' 'For heaven sake,' I cried' 'that was the celebrated caviar.' 'So *that* was caviar', he said, in wonderment. And after a short silence he added: 'Well, if you offer gourmet foods to peasants like me, you know they won't appreciate it'.[51]

In later years when Einstein was already old, sitting in Princeton and reminiscing, Einstein wrote to Solovine on 3 April 1953 from Princeton,[52]

"To the immortal Olympia Academy,

During your short active life of existence you took a childish delight in all that was clear and reasonable.[53]

Einstein went on,[54]

"Your members created you to amuse themselves at the expense of your big sisters who were older and puffed up with pride. I learned fully to appreciate just how far they had hit upon the true through careful observations lasting for many long years.

We three members, all of us at least remained steadfast. Though somewhat decrepit, we still follow the solitary path of our life by your pure and inspiring light; for you did not grow old and shapeless along with your members like a plant that goes to seed.

To you I swear fidelity and devotion until my last learned breath!"

Einstein had earlier written Solovine on November 25, 1948, that, they had wonderful time in those days in Berne in their cheerful "Academy", which was less childish than those respectable ones which he later got to know at close quarters. Einstein wrote Solovine that during the last months, one of Conrad Habicht's sons, a mathematician (like his father) came "here". Einstein heard from him news of the "old man" Habicht, and this reminded him that, "We really had wonderful time in Bern".[55]

## 2.3 The Reading List of the Academy

The reading list of the Academy included the following books and papers.[56]

Solovine and Einstein started to read Pearson's *Grammar of Science*, before Habicht joined them. After Habicht had joined the group, the three of them read and discussed Ernst Mach's *Analysis of Sensation* (*Analyse der Empfindunge*) (1900 or 1902 or 1903), and *Die Mechanik in ihrer Entwicklung* (*Science of Mechanics*, first edition 1893, but they probably read the fourth or fifth edition of 1901 or 1904).

They also read John Stuart Mill's *System of Logic*, David Hume's, *A Treatise of Human knowledge*, translated partly in 1895 as *Traktat über die menschliche Natur*, and Spinoza's Ethics, *Ethik*.

A few papers by Helmholz, and Reimann's 1854 Famous Göttingen lecture, "Über die Hypothesen welche der Geometrie zu Grunde liegen", "On the Hypotheses which lie at the Bases of Geometry" were also read and discussed.

Several chapters of *Aufsatz über die Philosophie der Wissenschaften* or the French edition *L'essai sur la philosophie des sciences* of 1834 by Ampère were also read.

---

William Kingdon Clifford's 1903 essay "On the Nature of Things-in-Themselves", and writings by Richard Dedekind's "Was Sind und was Sollen die Zahlen?" and Richard Avenarius.

Finally, the group also read Poincaré's 1902 La science et l'hypothese. Einstein may have used the German 1904 edition *Wissenschaft und Hypothese*. Solovine at an older age recalled that Poincaré's book *Science and hypothesis* "profoundly impressed us and kept us breathless for many weeks".[57] However, Solovine remembered that the academy members "devoted weeks to the discussion of David Hume's eminently penetrating criticism of conceptions of substance and causality". [58] Einstein often acknowledged his intellectual debt to Hume. For Einstein, Hume's philosophy complemented Mach's and Avenarius' works.

On the extensive reading list of the Olympia Akademie Kant's works were absent.[59]

In a letter of 14 April 1952 to Carl Seelig, Solovine includes Poincaré's 1905 *La valeur de la science*, in a less complete list of the readings of the Olympia Academy.[60] *La valeur de la science*, included his 1898 paper "The measurement of time" as chapter II, but in a revised form. In addition Poincaré's 1904 Saint Louis talk comprises chapters VII to IX of this book.[61] In the first paper Poincaré cited the synchronization of clocks by telegraphic signals for longitude measurements. In the second paper he gave a physical interpretation of the local time in terms of clock synchronization by light signals, and formulated a principle of relativity.

The book *La valeur de la science* was published in March 1905. Einstein probably did not read it before submitting his relativity paper in June 1905.

In 1904 Habicht left Bern to become a schoolteacher in his home town of Schaffhausen, where Einstein had once briefly tutored. Solovine, left a year later,

---

[57] "*La Science et l'hypothesèse* de Poincaré, un livre qui nous a profondément impressionnés et tenus en haleine pendant de longues semaines…". Einstein, 1956, p. VIII.
[58] "Nous avons discuté pendant des semaines sur la critique singulièrement sagace des notions de substance et de causalité faite par David Hume". Einstein, 1956, p. VIII; Sonnert, Gerhard, *Einstein & Culture*, 2005, New York: Humanity Books, p.305
[59] Sonnert, 2005, p.303; Einstein, 1956, pp. VII-VII.
Einstein read Kant as a boy with Talmud, and this was the first time he appreciated Kant. Toward 1918 Einstein appeared to become interested again in Kant's writings. In summer 1918 Einstein went on a summer vacation with his second wife Elsa. From there he sent Max Born a few letters. From an undated letter written to Born Einstein had written: "I am reading Kant's *Prolegomena* here, among other things, and am beginning to comprehend the enormous suggestive power that emanated from the fellow, and still does. […] Anyway it is very nice to read, even if it is not as good as his predecessor Hume's work." Einstein, Albert, *Albert Einstein Max Born Brief Wechsel 1916-1955*, 1969/1991, Austria: Nymphenburger, letter 5, 1918. Einstein told Carl Seelig: I did not grow up in the tradition of Kant". Seelig, 1956, p. 135; Seelig, 1954, p. 114. And Seelig told the following anecdote related to Kant: Einstein was asked after a lecture whether his theories did not conflict with Kant's philosophy. "That's difficult to say" he replied. "Each philosopher reads Kant in his own way". Seelig, 1956, p. 81; Seelig, 1954, p. 96.
[60] *CPAE*, vol 2, p. xxiv.
[61] Poincaré, Henri, *La valeur de la science*, 1905/1970 Paris: Flammarion; translated to English by F. Maitland, *The Value of Science*, New-York: Dover, pp. 41-54; pp. 123-147.

settled in Paris as an editor and writer, later becoming the authorized translator of Einstein's books into French. Thus the Olympian academy had only a brief existence, but the three friends kept in touch, and the academy lived in their memories;[62] and also in their letters…

This was the reason for why Poincaré's *La valeur de la science* could very likely not have actually been on the reading list of the Olympia Academy. Einstein could certainly read Poincaré's book *La valeur de la science*. But since he did not mention this book he either did not read it or he read it later.

### 3. *Annus mirabilis*

### 3.1 Letters to Habicht

By 1904 Einstein's close friend, Conrad Habicht, did not live any more in Bern. Einstein longed for the happy days of the Olympian Academy meetings; and to discuss his ideas with Habicht. On the first of August, 1904, Einstein wrote Habicht, "Your postcard just arrived. Please do come *at once*". I am still on vacation this week, and nothing would please me more than to see you here".[63] Five days later on Saturday August 6, Einstein wrote Habicht again, "Dear Habicht, You are an abominable creature. − − − All the rest hopefully soon in person in Bern. Regards to you and your family from your enraged E. S[olivine] left for Lyons".[64]

Einstein felt so lonely on Saturday, August 6, without Habicht, that he wrote him another letter that very day,

"Dear Habicht,

Did you only want to tease me, you abominable creature, or have you perhaps not received the card I sent you Monday? Out of rage and desperation, Solo [Solovine] left (for Lyons); he could not wait any longer for financial reasons.

Secretly I still hope to see you soon, even if only on Sunday, as vacation time has come to an end, unfortunately.

Cordially, your

E."[65]

Why was Einstein so in need of his friend in the beginning of August 1904? What did he want to tell him? May be he wanted to tell him about his 1905 "miraculous year" discoveries.

The next surviving letter is dated a few months later. On March 6 1905 Einstein again wrote Habicht two letters on the same day: "Don't forget Bern and the academy, and be sure to come and see us",[66] and: "You are herewith warned and challenged to attend a number of academic meetings of our praiseworthy 'Academy' and to increase forthwith by 50 per cent your present membership fee".[67]

Einstein very likely did not write Habicht between March and mid-May 1905. He was busily working during these two or three months on his *Annus mirabilis* papers. Two months after the last letter, Einstein sent Habicht another letter from Bern. The letters undated, but appear to have been written between May 18 and 25 1905:[68]

"Dear Habicht! A strange silence seems to reign between us, so that it appears to me to be almost a blasphemy if I now break it with a little insignificant prattle. But does this not always happen to the superior people of this world? What are you up to, you cold-blooded old whale, you dry bookish creature, or whatever I can still fling at your head, filled as I am with seventy per cent anger and thirty per cent compassion. You only have this last thirty per cent to thank that when recently you let Easter pass without a sound from you, I did not send you a tin of chopped onions and garlic. But why have you not yet sent me your thesis? Don't you know, you miserable, that I should be one of the few fellows who would read it with interest and pleasure? I can promise you in return four works, the first of which I shall soon be able to send you as I am getting some free copies. It deals with the radiation and energy characteristics of light and is very revolutionary, as you will see if you send me your work in advance. The second work is a determination of the true atomic dimensions from the diffusion and inner friction of diluted liquid solutions of neutral matter. The third proves that on the premise of the molecular theory of heat, particles of the size 1/1000 mm, when suspended in liquid, must execute a perceptible irregular movement which is generated by the movement of heat. Movements of small, lifeless, suspended particles have in fact been examined by physiologists, and these have been called by them 'Brownian molecular movement'. The fourth work is only a concept at this point, and is an electrodynamics of moving bodies, which employs a modification of the theory of space and time; the purely kinematical part of this work will surely interest you.

Solo (Solovine) spends much time giving lessons…

Greetings from your Albert Einstein. Greetings, too, from my wife and my year old chick."

Einstein would soon tell his friend Habicht about yet another fifth path breaking paper; an addendum to his work on the electrodynamics of moving bodies containing

---

[66] Einstein to Habicht, March 6, 1905, *CPAE*, Vol. 5, Doc. 25.
[67] Einstein to Habicht, March 6, 1905, *CPAE*, Vol. 5, Doc. 26; Seelig, 1956, p. 74; Seelig, 1954, p. 88.
[68] Einstein to Habicht, 18 or 25 May, 1905, *CPAE*, Vol. 5, Doc. 27; Seelig, 1956, pp. 74-75; Seelig, 1954, pp. 88-89.

Einstein's first derivation of the mass-energy equivalence (and in this letter persuade him to come and work with him in the Patent Office).

## 3.2 Einstein's *Annus mirabilis* Papers

In 1905, Einstein was twenty-six-years-old. He sat in the Patent Office in the building of the postal and Telegraph Administration. He probably sent his papers to the *Annalen der Physisik* from there. Einstein signed, "Bern, June, 1905", in account with the *Annalen* regulations, he wrote the name of the city from which the paper was sent. Many scholars would look for him at the University of Bern, as no one would think to go to the Patent Office.[69]

Einstein's burst of ideas in the *Annus mirabilis* of 1905 was amazing. He published five papers in the *Annelen der Physik* that solved some of the most persisting problems confronting the *fin de siècle* physics of the turn of the 19[th] and 20[th] centuries. Of course prior to publication of these papers Einstein was immersed in years of tedious work which led to this burst of creativity:[70]

1) "On a Heuristic Viewpoint Concerning the Generation and Transformation of Light", sent to the *Annalen* on March 17[th], 1905, and received by the *Annalen* a day afterwards (three days after Einstein 26[th] Birthday). It argues a heuristic manner for the existence of light quanta and derives the photoelectric law. Einstein won the 1922 Nobel Prize for physics ostensibly for this work.

2) "On a New Determination of Molecular Dimensions", on his doctoral thesis submitted to the mathematical and natural science branch of Zürich University. (In the Polytechnic the first doctorate degree could only follow in the autumn of 1909). The thesis is dedicated "to my friend Dr. Marcel Grossmann" and was approved by Einstein's mentor Professor Alfred Kleiner and Professor Heinrich Burkhardt.

The elder Joseph Sauter told Seelig in the 1950s that, he told Einstein to send his doctoral thesis to Zürich, because "it would be absolutely child's play for you to get a doctorate there".[71] Einstein's sister Maja reported in 1924 that, at first Einstein attempted to submit his recently completed paper, "On the Electrodynamics of Moving Bodies" (the theory of relativity) as a Doctoral thesis to the University of Zürich: "but the thing didn't seem quite right to the leading professors, as the wholly unknown author paid no heed to authority figures, even attacked them! So the work was simply rejected (irony of fate!) and the candidate saw himself compelled to write and submit another, more harmless work, on the basis of which he then obtained the title of Doctor Philosophaie"[72] In 1907 Einstein applied for a teaching position at the University of Bern. He again failed to follow the rules: he was supposed to send along with the application a not hitherto published scientific paper, and he enclosed the 1905 relativity paper (see section 5.1).

[69] Winteler-Einstein, 1924, p. 24.
[70] Seelig, 1956, p. 68; Seelig, 1954, pp. 81-82.
[71] Seelig, 1956, p. 73; Seelig, 1954, p. 87.
[72] Winteler-Einstein, 1924, p. 23.

3) "On the Movement of Particles Suspended in Fluids at Rest, as Postulated by the Molecular Theory of Heat". This is the Brownian motion paper. It was received by the *Annalen* on May 11, 1905.

4) "On the Electrodynamics of Moving Bodies", received by the *Annalen* on June 30, 1905. The relativity paper, thirty printed pages, the manuscript of which was destroyed after publication.

5) "Does the Inertia of a Body Depend on its Energy Content?" The first derivation of the mass energy equivalence; received by the *Annalen* on September 27, 1905.

Dukas and Hoffmann say, "He probably did his calculations on odd slips of paper, and although in submitting his now-famous works of 1905 to *Annalen der Physik* he presumably wrote them out reasonably neatly, once they were in print he discarded the manuscripts, perhaps after using them as scraps of paper on the backs of which to perform other calculations. Thus the originals no longer exist. But that was the way Einstein was".[73] Hoffmann and Dukas add, "Einstein had completed the doctoral thesis and the paper on the Brownian motion – all while earning his living full-time at the Patent Office. No wonder he felt exhausted when the relativity paper was done".[74]

Einstein was now Herr Doktor and hence Friedrich Haller addressed to the Swiss Federal Council, proposing Einstein's long overdue promotion. Haller wrote that Einstein had "increasingly familiarized himself with technology, so that he now very successfully processes quite different technical patent applications and is one of the most highly respected experts of the Office".[75]  On April 1, 1906, Einstein became an Expert II Class. Until then Einstein's initial salary was 3500 francs annually. Now his salary went up by 600 francs to 4500 francs annually.[76] When Haller informed him of his promotion, says Seelig, a question was put which is seldom heard in a civil servants' office: "But what shall I do with all that money?"[77]

## 4 German Scientists Respond to Einstein's Relativity Paper

### 4.1 Professor Max Planck Writes Einstein

Maja reports that when Einstein submitted his work on relativity in the year 1905 to the *Annalen de Physik*, he was afraid for some time that the work might be rejected.[78]

Once his paper was published he expected sharp and severe criticism. The *Annalen* was widely read by young scholars, and thus Einstein expected immediate reaction to it. However, he was very disappointed, because complete silence followed. The following issues did not mention his publication and he received no other reaction to the paper. The professionals were silent for some time after the first appearance of the paper.[79]

Finally Einstein received a letter from Berlin. It came from Professor Max Planck, the well-known physicist who asked him to explain some obscure points in the paper. After Einstein long wait this was the first sign, according to Maja that, his work had been actually read. "The joy of the young scholar was so great", because he received recognition from one of the greatest living physicists.[80]

Maja reports that in the following years Einstein and Planck corresponded and despite the age difference a solid friendship developed between them. Planck raised Einstein's morale. According to Maja he brought to light the genius of the patent clerk by spreading the theory and organizing a seminar on the new theory with his students. These also became interested in what seemed to be unsolved problems of the theory, and started to correspond with the young physicist in Bern.[81] Naturally, says Maja, they sent their letters to "Professor Einstein" and directed them to the University of Bern, "because no one suspected that the author of the publications that by then had aroused great stir had a humble official position in the Patent Office".[82]

Einstein had to write Planck twice before Planck answered him on 6 July 1907: "Your postcard of the 3[rd] of this month reminds me that I still owe you an answer to your valued letter of 6 June, which I had already long ago resolved to give you".[83] In the letter of July 6, 1907 Plank wrote Einstein, "Returning to a remark in one of your last letters, I would like to add that I shall probably be going to the Bernese Oberland next year. Admittedly, this is rather far in the future, but I am happy to think that perhaps I shall then have the pleasure of making your personal acquaintance".[84] However Einstein and Planck only met two years later in the Naturforscherversammlung in Salzburg in 1909, the first nationwide congress attended by Einstein.

## 4.2 Max Laue Meets Einstein

Planck had an assistant in Berlin, appointed in the autumn of 1905, Max Laue, born in the same year as Einstein.[85] Laue wrote Einstein and asked to meet him in Bern in the summer of 1906. It seems – though the evidence is unclear say Hoffmann and Dukas – that Laue automatically assumed that Einstein was at Bern University (like the other scholars according to Maja's above report) and went there to look for Einstein.

Seelig tells the story as Laue recounted it. Laue used his summer holiday, after a mountain tour, to become personally acquainted with the creator of the theory of relativity in Bern. Of the meeting he told Seelig:[86]

"After a written appointment I looked him up in the Patent Office. In the general waiting room an official said to me: 'Follow the corridor and Einstein will come out and meet you.' I followed his instructions but the young man who came to meet me made so unexpected an impression on me that I did not believe he could possibly be the father of the relativity theory. So I let him pass and only when he returned from the waiting room did we finally become acquainted. I cannot remember the actual details of what we discussed but I do remember that a cigar he offered me was so unpleasant that I 'accidently' dropped it into the river from the Aare bridge. […] In any case from that visit I came away with some understanding of the relativity theory".

Planck wrote Einstein on November 9, 1907, "Thank you so much for your kind letter and the news, which was of great interest to me. Herr Laue had also told me about his very pleasant meeting with you".[87]

## 5. Einstein Teaches his 3 Friends at the University of Bern

## 5.1 Patent Clerk Rebels against Academic Rules

Shortly after Einstein started work at the Patent Office, in 1902, Sauter took Einstein to the meetings of the *Naturforschende Gesellschaft* (Natural Science Society) in Bern, an association of professors, high school teachers and prominent figures in medicine and pharmacology, of which Sauter was a regular member. On May 2, 1903 Einstein met there a friend of Sauter's, Dr. Paul Gruner, at that time a high school

---

teacher and simultaneously a privatdozent in physics at Bern University.[88] In 1906 a professorship of theoretical physics was created at the University of Bern and the post was given to the 54-year-old Gruner. [89]

In 1907 Einstein decided too to apply for a post of a privatdozent while retaining his position at the Patent Office. Abraham Pais writes that, it used to be said often in those times that a university career could be contemplated only if one was independent, wealthy or married to a well-to-do person; of course, neither applied to Einstein.[90] Seelig writes that, Einstein consulted with Gruner, who had learned of his extraordinary capabilities, and then in 1907 he applied for admission to the faculty of theoretical physics. [91]

On June 17, 1907 Einstein sent a letter to the cantonal authorities in Bern enclosing copies of his doctorate thesis, of seventeen published papers (including of course the path breaking ones of 1905), and a curriculum vitae. Several faculty members spoke in favor of the application when the matter came up for discussion. Einstein (the rebellious) failed to follow the rules and had omitted to follow the requirement to send along with his application the *Habilitationsschrift*, *habilitation* – a not hitherto published scientific paper.[92] He enclosed as the *habilitation* the 1905 relativity paper.[93]

There was a Professor Aimé Forster, who had been teaching at Bern University since 1896. Forster, the professor for experimental physics, took the view that it was unnecessary and unsuitable to engage, in addition to Gruner, a privatdozent in the field of theoretical physics in which there were no more than half a dozen students. He returned Einstein's work, "The Electrodynamics of Moving Bodies" with the following remark: "I can't understand a word of what you've written here!". [94] He could not easily accept a work which attacked the tacit assumptions of physics, and he expected a work that followed the accepted rules. In fact when applying to a position,

Einstein should have submitted a work according to the rules in light of the faculty members in the department (Aimé Forster).

Einstein was so irritated that he momentarily abandoned the idea of pursuing an academic career. Einstein's request was denied until such time as Herr Einstein saw fit to produce the *habilitation*.

Meanwhile, Einstein's close friend from the Polytechnic, Marcel Grossmann, had become professor at the Zürich Polytechnic in 1907, succeeding Professor Otto Wilhelm Fiedler. Grossman was not like his close friend Herr Einstein, a rebel, vagabond and Eigenbrödler (loner)[95] On January 3, 1908 despairing and irritated Einstein wrote to Grossman, asking him the best way to apply for a vacant high school position:[96]

"At the risk of your thinking me ridiculous, I must ask your advice on a practical matter. I want very much to launch an attack on a teaching position at the Technical School in Winterthur (Mathematics and Physics). A friend who is a teacher there has told me in strictest confidence that the position will probably become vacant pretty soon. […] I come to this hankering only because of an ardent desire to be able to continue my private scientific work under less unfavorable conditions, as you will certainly understand. […] I once taught there for a few months as a substitute teacher. […] I now ask you: How does one go about it? Should I perhaps pay someone a visit so as to demonstrate to him face to face the high value of my worthy self as teacher and citizen? Who would be the man to see? Would I not be likely to make a bad impression on him (not speaking Swiss German, Semitic features, etc)? Furthermore, would it make sense if, at this interview, I were to extol my scientific papers?"[97]

Later, Einstein changed his mind about the *Habilitation* thesis. Einstein wrote Gruner on February 11, 1908: "The conversation that I had with you in the Library, as well as the advice of several friends has caused me to change my mind again and to make a further application for admission to Berne University. To this end I have forwarded a *habilitation* to the Dean".[98]

Hoffman and Dukas write that Einstein changed his mind because, "Professor Alfred Kleiner – he who had been involved in both the rejection and acceptance of doctoral theses submitted by Einstein to Zürich University – sent Einstein a postcard expressing a desire to get in touch with him about a matter of mutual importance.[99] Seeking to bring Einstein to Zürich University, Kleiner urged him not only to try once more to become *Privatdozent* at Bern University but also to report any developments, so that if things went badly, Kleiner could try to think of less orthodox ways in which Einstein might meet the prerequisites for a professorship:[100]

---

[95] Einstein, Albert, "Erinnerungen-Souvenirs", *Schweizerische Hochschulzeitung* 28, Sonderheft, 1955, pp. 145-148, pp. 151-153; Reprinted as, "Autobiographische Skizze" in Seelig Carl, *Helle Zeit – Dunkle Zeit. In memoriam Albert Einstein*, 1956, Zürich: Branschweig: Friedr. Vieweg Sohn/Europa, pp. 9-17; p. 11.
[96] Einstein to Grossman, January 3, 1908, *CPAE*, Vol. 5, Doc. 71; Hoffmann and Dukas, 1973, p. 86.
[97] The job never opened up. The principal of the school at Winterthur wrote Einstein, "Unfortunately, the chances that a position will become vacant here very soon look dim at the moment". Adolf Gasser to Einstein, mid-January, 1908, *CPAE*, Vol. 5, Doc. 74.
[98] Einstein to Gruner, February 11, 1908, *CPAE*, Vol. 5, Doc. 81.

"You did this very promptly, this matter with the *habilitation*; but it seems to me that one must continue to keep an eye on the progress of this matter lest it get stuck somewhere. If, for example, Forster has to evaluate this work, he might easily procrastinate with the delivery of his vote if, for some reason or other, this *Habilitation* is not to his liking; I think therefore that you should try to interest some acquaintance [like Gruner] on the faculty in handling the agenda this semester." Kleiner wrote that he "must be kept informed about this matter, for if necessary, i.e., if you did not get to lecture next semester via *habilitation*, I must see to it that you be given an opportunity of another kind to prove yourself as a lecturer. (A lecturer at the *Naturforschende Ges* [ellschaft] or something similar). Please keep me up to date".

Kleiner then told Einstein to contact Gruner regarding his *habilitation*. Einstein prepared a new *Habilitation* thesis and submitted it to the Philosophical Department II.[101] The thesis has been circulated among the faculty members, and Professor Forster proposed that the thesis be accepted and that Herr Einstein be invited to an Inaugural lecture. On February 27, 1908 this lecture was given, "Consequences for the Constitution of Radiation of the Energy Distribution Law of Black Body Radiation",[102] and the faculty members recommended that the candidate be made a Privatdozent for theoretical physics.[103]

## 5.2 Einstein's Students: Besso and Chavan

Einstein gave his Bern lectures in the winter term 1908-1909 in the evening from six to seven p.m. on the theory of radiation (the subject of his *Habilitation* thesis).[104] Einstein's first audience consisted of his two friends from the Patent Office, Michele Besso and Lucian Chavan, and another colleague from the Patent Office, Heinrich Schenk. Chavan reported on these lectures in his notebook (written in French) that was deposited in the Federal Postal and Telegraph Administration library after his death in August 1942. Flückiger's book reproduces two pages of these notebooks, and

a description: Meticulous, clean, and ordered pages of the notebooks, with figures and equations.[105]

In the summer term, Einstein started with the course "Kinetic Theory of Heat". On Tuesday and Saturday of each week, Einstein, Besso and Chavan had to get up early and climb up the Grosse Schanze, where Einstein would begin the class at seven in the morning so that he and his colleagues could start work at the Patent Office at eight. In 1909 Besso and Chavan gave up and Einstein was left only with one student Max Stern. For this reason, he cancelled his class and sent Stern a postcard saying that he would not deliver the lectures but that he was always prepared to give him advice outside the university.[106]

On the first of March, 1908 Jacob Johann Laub (an Austrian-born, who graduated under Professor Wilhelm Wien, professor of physics at the University of Würzburg) wrote Einstein, "I must confess to you that I was surprised to read that you have to sit in an office for eight hours a day. But history is full of bad jokes".[107]

### 5.3 Einstein Leaves the Patent Office to his First Post in Zürich

In the winter semester of 1908-1909 the University of Zürich looked for a professor of physics. Professor Alfred Kleiner already had a candidate for the post – Friedrich Adler, Einstein's acquaintance from his student days in the Polytechnic of Zürich. Adler in the meantime had found an interesting job at the German Museum in Munich. But Kleiner persuaded Adler to return to Zürich. Adler was the natural candidate, but Kleiner decided to consider another scientist. Adler suggested Einstein.

Adler wrote his father on 19 June, 1908 that his competitor "is a man named Einstein, who studied at the same time as I did. We even heard a few lectures together. Our development is seemingly parallel: He married a student at about the same [time] as I, and has children. But no one supported him, and for a time he half starved. As a student he was treated contemptuously by the professors, the library was often closed to him, etc. He had no understanding of how to get on with important people. Finally he found a position in the Patent Office in Bern and throughout the period he has been continuing his theoretical work in spite all distractions. Today he is in the school of Boltzmann, and one of the most distinguished and recognized, and this school, not that of Mach is the mode today."[108]

At the beginning of 1909, while Einstein was a *Privatdozent* at the University of Bern, Alfred Kleiner attended one of Einstein's lectures. His impression was that Einstein lectured poorly. Einstein later reported about this to Laub, [109]

"When you wrote to me that the news that I lecture poorly had reached you via [Matthias] Cantor, I read this negative, because, I immediately knew the route by which this rumor traveled: Kleiner – Burkhard – Cantor – Laub – Einstein. For the former [Kleiner], once came to my class to inspect the beast. On that occasion I really did not lecture divinely – partly because I was not prepared very well, partly because the state of having-to-be-investigated got on my nerves. I then wrote a letter to Kleiner, in which I seriously reproached him for spreading unfavorable rumors about me and thus turning my position, which is already so difficult, into a final and definitive one. For such a rumor destroys any hope of getting into the teaching profession. K. then repented. He said that he would be pleased to get me an extraordinary professorship in Zürich if he could satisfy himself that I have some teaching ability. Then I suggested to him a lecture at the Physik. Gesellschaft Zürich, which I delivered 3 months ago. I was lucky. Contrary to my habit, I lectured well on that occasion – and so it has come to pass".

Einstein did not expect much, and said: "I certainly don't demand to be made a university professor at Zürich".[110] Einstein gave a lecture to the "Physical Society", and there he was able to correct the impression. On April 28th, 1909 Einstein reported to his close friend Conrad Habicht: "I am fairly sure now of getting my post at Zürich University".[111]

The faculty was not eager to accept Einstein. They wrote, "Herr Dr Einstein is an Israelite and since precisely to the Israelites among scholars are ascribed (in numerous cases not entirely without cause) all kinds of unpleasant peculiarities of character, such as intrusiveness, impudence, and a shopkeeper's mentality in the perception of their academic position. It should be said, however, that also among the Israelites there exist men who do not exhibit a trace of these disagreeable qualities and that it is not proper, therefore to disqualify a man only because he happens to be a Jew". And the committee and faculty did not consider it compatible with their dignity in democratic Zürich to adopt anti-Semitism as a matter of policy; so that the information which Kleiner provided about Einstein "reassured" them.[112]

On May 7, 1909 Einstein was appointed by the Governmental Council of the Canton of Zürich, for the period of six years, with a salary of 4500 francs; and he was to assume his position at the beginning of the winter semester on October 15, 1909. He wrote Laub in the same letter quoted above, "So now I am an official of the guild of whores, etc".[113] Einstein called Kleiner-Burkhard-Cantor and their Colleagues, "der Gilde der Huren", because they spread unfavorable rumors about him, while they knew that he had difficulties obtaining the position in Zürich. Einstein felt uncomfortable to be an official of such a "Gilde" of people, and he said that about university professors in general.

On July 6, 1909, Einstein handed his resignation notice from the Patent Office to Haller, effective October 15 1909.[114] Haller placed on record that the Expert II Class had performed highly valued services. His departure is a loss to the office. However, Herr Einstein feels that teaching and scientific research are his real profession, and for that reason the Director of the Office made no attempt to bind him to the Office by better financial arrangements.[115]

*I wish to thank Prof. John Stachel from the Center for Einstein Studies in Boston University for sitting with me for many hours discussing special relativity and its history. Almost every day, John came with notes on my draft manuscript, directed me to books in his Einstein collection, and gave me copies of his papers on Einstein, which I read with great interest.*